\documentstyle{article}[12pt]
\newcommand{\nin}{\noindent}
\newcommand{\be}{\begin{equation}}
\newcommand{\ee}{\end{equation}}
\newcommand{\bea}{\begin{eqnarray}}
\newcommand{\eea}{\end{eqnarray}}

\newcommand{\hf}{\frac{1}{2}}
\newcommand{\nonu}{\nonumber\\}

\begin{document}

\begin{center}

{\Large{\bf Non-renormalization\\ for planar Wess-Zumino model}}

\vspace{.5cm}

{\bf Jean Alexandre}\\
Physics Department, King's College \\
London, WC2R 2LS, UK\\
jean.alexandre@kcl.ac.uk

\vspace{1cm}

{\bf Abstract}

\end{center}

Using a non-perturbative functional method, where the quantum fluctuations are gradually set up,
it is shown that the interaction of a $N=1$ Wess-Zumino model in 2+1 dimensions does not
get renormalized. This result is valid in the framework of the gradient expansion and
aims at compensating the lack of non-renormalization theorems.

\vspace{2cm}

Non-renormalization theorems are based on analyticity properties and are thus not 
present for $N=1$ supersymmetry in 2+1 dimensions, since the odd coordinate of superspace
is real. Supersymmetric properties should nevertheless restrict the quantum corrections
in this situation, and lead to a control of the renormalization processes.

In the framework of relativistic-like effective descriptions of high-temperature superconductivity
\cite{highTc}, supersymmetric models in 2+1 dimensions have been introduced \cite{susy}, 
where conditions for the elevation to a $N=2$ supersymmetry were studied, motivated by the presence 
of non-renormalization theorems in this case. It is though interesting to look at the possibility 
to have exact results with $N=1$ supersymmetry and a functional method is presented here 
for this purpose, which 
gives indications on the renormalized structure of a Wess-Zumino model in 2+1 dimensions.

The idea of the method is to control quantum fluctuations with the mass of the bare theory.
When this mass is very large, the quantum fluctuations are frozen and the system is classical.
As the bare mass decreases, the quantum fluctuations appear and the parameters of the theory
get dressed. We can then consider the "fluctuation flows" of the effective action 
(the proper graphs generator functional) with the bare mass  
and, starting from the bare action, follow these flows so as to built the "full" quantum theory,
containing all the quantum effects.

There are similarities between this procedure and the blocking procedure \cite{block}, since the 
latter describes the evolution of a theory with a momentum, from the ultraviolet (UV) scales to
the infrared (IR) ones. The present method, though, is not based on a splitting the UV from the IR 
degrees of freedom and thus does not introduce any artificial coarse graining function. 
It was shown, in previous works \cite{int,control}, that these fluctuation flows recover the 
usual one-loop results. Beyond one-loop, the results given by these flows do not co\"{\i}ncide anymore with
a loop expansion, since the results are based on the so-called gradient expansion.
Note finally that this method does not require any regularization in 2+1 dimensions, which is 
another advantage to use it. 

We will find, in the context of the gradient expansion, the exact effective action for 
a $Q^4$ Wess-Zumino theory ($Q$ is the scalar superfield and $Q^4$ leads to the marginal
interaction $\phi^6$, where $\phi$ is the scalar component of $Q$). We will see that only
the mass gets renormalized, whereas the interaction does not. 

\vspace{.5cm}

We note $z=(x,\theta)$ the coordinate of superspace and we take the conventions used in \cite{gates}.
The bare action is 

\be\label{bareaction}
S_\lambda[Q]=\int d^5z\left\{\hf Q D^2 Q+\frac{\lambda}{2}m_B Q^2+\frac{g_B}{24}Q^4\right\},
\ee

\nin where $g_B$ is the dimensionless bare coupling. The effective action $\Gamma_\lambda$,
defined as the Legendre transform of the connected graphs generator functional, depends on 
$\lambda$: for $\lambda>>1$ $\Gamma_\lambda$ describes the classical theory, and as $\lambda$
decreases down to 1 the quantum theory emerges out of the increasing quantum fluctuations. 
The massless case can be obtained by letting $\lambda$ decrease down to 0.

The exact and non-perturbative evolution equation of the effective action with the 
parameter $\lambda$ was derived in \cite{control} and reads

\be\label{evol}
\partial_\lambda\Gamma_\lambda=\frac{m_B}{2}\mbox{Tr}\left\{Q^2
+\left(\frac{\delta^2\Gamma_\lambda}{\delta Q\delta Q}\right)^{-1}\right\},
\ee

\nin where the trace "Tr" is to be taken over the superspace coordinates. 
In the framework of the gradient expansion, we consider the following ansatz for the
functional dependence of $\Gamma$:

\be\label{gradexp}
\Gamma_\lambda[Q]=\int d^5z\left\{\hf Q D^2 Q+U_\lambda(Q)\right\},
\ee

\nin i.e. we allow any potential term, but do not take into account higher
order kinetic terms or derivative interactions. In this situation, in order to find the evolution
of the potential, it is enough to consider a
constant configuration $Q_0$ of the superfield $Q$. $Q_0$ is a vacuum expectation value of
the scalar component of $Q$. We have 
in principle $\int dp^3=\infty$ and $\int d^2\theta=0$, but we consider the 
finite, regularized, volume of superspace

\be
{\cal V}=\int\frac{dp^3}{(3\pi)^3}d^2\theta.
\ee
 
\nin We have then for the effective action

\be
\Gamma_\lambda[Q_0]={\cal V}U_\lambda(Q_0),
\ee

\nin and for its second derivative

\bea
&&\frac{\delta^2\Gamma_\lambda}{\delta Q(p,\theta)\delta Q(q,\theta')}|_{Q=Q_0}\nonu
&=&\left[U_\lambda^{(2)}(Q_0)+D^2_{p\theta}\right](2\pi)^3\delta^3(p+q)\delta^2(\theta-\theta'),
\eea

\nin where $U_\lambda^{(2)}(Q_0)$ denotes the second derivative of the potential with respect to $Q_0$.
Using then the properties \cite{gates}

\bea\label{propgates}
&&\delta^2(\theta-\theta')\delta^2(\theta'-\theta)=0\nonu
&&\delta^2(\theta-\theta')D^2_{p\theta}\delta^2(\theta'-\theta)=\delta^2(\theta-\theta')\nonu
&&(D^2_{p\theta})^2=-p^2,
\eea

\nin we easily obtain 

\bea
&&\left(\frac{\delta^2\Gamma_\lambda}{\delta Q(p,\theta)\delta Q(q,\theta')}|_{Q=Q_0}\right)^{-1}\nonu
&=&\frac{U_\lambda^{(2)}(Q_0)-D^2_{p\theta}}{p^2+[U_\lambda^{(2)}(Q_0)]^2}
(2\pi)^3\delta^3(p+q)\delta^2(\theta-\theta'),
\eea

\nin and then

\bea
&&\mbox{Tr}\left(\frac{\delta^2\Gamma_\lambda}{\delta Q\delta Q}|_{Q=Q_0}\right)^{-1}\nonu
&=&\int\frac{dp^3}{(2\pi)^3}\frac{dq^3}{(2\pi)^3}d^2\theta d^2\theta'\delta^3(p+q)\delta^2(\theta-\theta')\nonu
&&~~~~\times\left(\frac{\delta^2\Gamma_\lambda}{\delta Q(p,\theta)\delta Q(q,\theta')}|_{Q=Q_0}\right)^{-1}\nonu
&=&-{\cal V}\int\frac{dp^3}{(2\pi)^3}\frac{1}{p^2+\left[U_\lambda^{(2)}(Q_0)\right]^2}.
\eea

\nin This last integral is divergent but can be written:

\bea
&&\mbox{Tr}\left(\frac{\delta^2\Gamma_\lambda}{\delta Q\delta Q}|_{Q=Q_0}\right)^{-1}\nonu
&=&-\frac{{\cal V}}{2\pi^2}\int_0^\infty dp
+\frac{{\cal V}}{2\pi^2}\int_0^\infty dp \frac{[U_\lambda^{(2)}(Q_0)]^2}
{p^2+[U_\lambda^{(2)}(Q_0)]^2}\nonu
&=&\mbox{Const.}+\frac{{\cal V}}{4\pi}U_\lambda^{(2)}(Q_0),
\eea
 
\nin where the diverging constant does not depend on $Q_0$.
Plugging this result in the evolution equation (\ref{evol}) and discarding the field-independent
terms, we finally obtain 

\be\label{finalequa}
\partial_\lambda U_\lambda(Q_0)=\frac{m_B}{2} Q_0^2+\frac{m_B}{8\pi}U_\lambda^{(2)}(Q_0).
\ee

\vspace{.5cm}

Let us now turn to the solution of this equation. Starting from the initial interaction (\ref{bareaction}),
we see that, when decreasing the fluctuation parameter from $\lambda$ to $\lambda-\delta\lambda$, the potential 
$U_{\lambda-\delta\lambda}$ will not acquire higher powers of $Q_0$ than the ones contained in $U_\lambda$ since

\be
U_{\lambda-\delta\lambda}(Q_0)=U_\lambda(Q_0)
-\delta\lambda\frac{m_B}{2}\left[Q_0^2+\frac{1}{4\pi}U_\lambda^{(2)}(Q_0)\right]
\ee

\nin As a consequence, we consider the following ansatz for $U_\lambda(Q_0)$:

\be
U_\lambda(Q_0)=u_0(\lambda)+\hf u_1(\lambda)Q_0^2+\frac{1}{24}u_2(\lambda)Q_0^4,
\ee

\nin and plug it in the evolution equation (\ref{finalequa}), to find 
the exact solution

\bea
u_2(\lambda)&=&g\nonu
u_1(\lambda)&=&\lambda m_B\left(1+\frac{g}{8\pi}\right)+M\nonu
u_0(\lambda)&=&\frac{m_B}{8\pi}\left[\frac{\lambda^2}{2}m_B\left(1+\frac{g}{8\pi}\right)+M\lambda+a\right],
\eea

\nin where $(g,M,a)$ are constants of integration to be determined with initial conditions.
To determine $g$, we invoke the central idea of this method, i.e. the fact that for $\lambda\to\infty$,
the theory should be the classical one and thus $g=g_B$. The choice of $M$ must be made with another 
boundary condition, since $u_1(\lambda)$ diverges as $\lambda\to\infty$. With an 
initial condition such that $\lambda>>1$, we should take a {\it finite} value of $\lambda$, and therefore
also an initial effective action which would already contain some quantum effects. To avoid this,   
we take the boundary condition 
at $\lambda=0$, where the bare theory is massless. There the bare action S does not contain any dimensionfull 
parameter, and also no mass parameter is introduced for any regularization, such that no dynamical 
mass can be generated in the quantum theory. We conclude that $M=0$. Finally, the constant $a$ is not 
important since it deals with field-independent terms.

Discarding the field-independent terms, the $\lambda$-dependent effective theory is eventually described by the 
following effective action

\be
\Gamma_\lambda[Q]=\int d^5z\left\{\hf QD^2Q+\frac{\lambda}{2}m_B\left(1+\frac{g_B}{8\pi}\right)Q^2
+\frac{g_B}{24}Q^4\right\}.
\ee

\vspace{.5cm}

To conclude, we see that no new interaction is generated by the quantum fluctuations and 
the bare interaction does not get any quantum correction, but only the
mass term does. The full quantum theory is then described by the effective action 

\be
\Gamma_{\lambda=1}[Q]=\int d^5z\left\{\hf QD^2Q+\frac{m_B}{2}\left(1+\frac{g_B}{8\pi}\right)Q^2
+\frac{g_B}{24}Q^4\right\},
\ee

\nin which is exact in the framework of the gradient expansion (\ref{gradexp}). This is consistent
with \cite{control}, where the potential was truncated to the order of bare interaction. 

We also see that the massless theory $(\lambda\to 0)$ does not get any quantum correction 
at all, in the approximation (\ref{gradexp}), since

\be
\Gamma_{\lambda=0}[Q]=\int d^5z\left\{\hf QD^2Q+\frac{g_B}{24}Q^4\right\}=S_{\lambda=0}[Q].
\ee

\nin This result should not be confused, yet, with a non-renormalization theorem: 
higher order derivative terms or derivative interactions would lead to a renormalization
of the potential and this study is let for a future work. But, as expected, 
the supersymmetric structure of the theory can still lead to strong predictions concerning
the quantum corrections.


\begin{thebibliography}{99}

\bibitem{highTc}
N.Dorey, N.Mavromatos, Nucl.Phys.B386 (1992): 614;
X.G.Wen, P.Lee, Phys.Rev.Lett.76 (1996) 503;
L.Balent, M.Fisher, C.Nayak, Phys.Rev.B60 (1999) 1654;
M.Franz, Z.Tesanovic, Phys.Rev.Lett.87 (2001) 257003;
I.Herbut, Phys.Rev.Lett.88 (2002) 047006;
see also references therein.

\bibitem{susy}
J.Alexandre, N.Mavromatos, S.Sarkar, Int.J.Mod.Phys.B17 (2003): 2359.

\bibitem{block}
F.Wegner, A.Houghton, Phys.Rev.A8 (1973) 40;
J.Polchinsky, Nucl.Phys.B231 (1984) 269;
C.Wetterich, Phys.Lett.B301 (1993) 90;
M.Reuter, C.Wetterich, Nucl.Phys.B391 (1993) 147;
T.Morris, Int.J.Mod.PhysA9 (1994) 2411;
U.Ellwanger, Phys.Lett.B335  (1994) 364;
N.Tetradis, D.Litim, Nucl.Phys.B464 (1996) 492;
J.Alexandre, V.Branchina, J.Polonyi, Phys.Lett.B445 (1999) 351.
see also references therein.

\bibitem{int}
J.Alexandre, J.Polonyi, Ann.Phys.288 (2001) 37;
J.Alexandre, J.Polonyi, K.Sailer, Phys.Lett.B531 (2002) 316;
S.Correia, J.Polonyi, J.Richert, Ann.Phys.296 (2002) 214;
J.Alexandre, hep-th/0310093.

\bibitem{control}
J.Alexandre, Phys.Rev.D68 (2003): 085016. 

\bibitem{gates}
S.J.Gates, M.Rocek, W.Siegel "Superspace", Benjamin/Cummings, Reading (1983).


\end{thebibliography}
\end{document}